\documentclass[11pt,a4paper]{article}
\usepackage[a4paper,total={160mm,220mm}]{geometry}

\usepackage{amsmath,amssymb}
\usepackage{graphicx}
\usepackage{cite}
\usepackage{caption}
\usepackage{subcaption}
\graphicspath{{img/}}

\usepackage{braket}
\usepackage{hyperref}

\newcommand{\abs}[1]{\left\lvert{#1}\right\rvert}

\newcommand{\md}{\mathrm{d}}
\newcommand{\me}{\mathrm{e}}
\newcommand{\phiq}{\tilde{\varphi}}
\newcommand{\op}{\hat{\mathcal{O}}}

\renewcommand{\vec}[1]{\mathbf{#1}}

\begin{document}
\numberwithin{equation}{section}
\title{
\vspace*{-0.5cm}{\scriptsize \mbox{}\hfill IPPP/22/09}\\
\vspace{3.5cm}
\Large{\textbf{Multiparticle Amplitudes in a Scalar EFT}
\vspace{0.5cm}}}

\author{Valentin V.~Khoze and Sebastian Schenk\\[2ex]
\small{\em Institute for Particle Physics Phenomenology, Durham University,} \\
\small{\em South Road, Durham DH1 3LE, United Kingdom}\\[0.8ex]}

\date{}
\maketitle

\begin{abstract}
\noindent
At sufficiently high energies the production of a very large number of particles is kinematically allowed.
However, it is well-known that already in the simplest case of a weakly-coupled massive $\lambda \varphi^4$ theory, $n$-particle amplitudes become non-perturbative in the limit where $n$ scales with energy.
In this case, the effective expansion parameter, $\lambda n$, is no longer small and the perturbative approach breaks down.
In general, the associated $n$-particle production rates were argued to be described by an exponential that, depending on the specifics of the underlying Quantum Field Theory model, could be either growing or decaying in the large-$n$ regime.
We investigate such processes in general settings of Effective Field Theory (EFT), involving arbitrary higher-dimensional operators of $\varphi$.
We perform the resummation of all leading loop corrections arising from EFT vertices for amplitudes at the multiparticle threshold.
We find that the net effect of higher-dimensional operators amounts to an exponentially growing factor.
We show that if an exponential growth was already generated by the renormalizable interactions, it would then be further enhanced by the EFT contributions.
On the other hand, if the multiparticle rates computed in the renormalizable part of the theory were suppressed, this suppression would not be lifted in the EFT.
\end{abstract}

\newpage

\section{Introduction}
\label{sec:introduction}

The most simple models of a Quantum Field Theory (QFT) are built of scalar degrees of freedom.
While their simplicity renders them an ideal laboratory for studying the dynamics of quantum fields at arbitrary energies, scalar QFTs typically suffer from fundamental shortcomings.
Perhaps, the most prominent examples of the latter are hierarchy as well as triviality problems (for reviews, see, e.g.,~\cite{Giudice:2008bi,Callaway:1988ya}).
In addition, scalar QFTs are afflicted by evidence of rapidly increasing rates of high-energy processes involving the production of a large number of particles.
Perturbative as well as semiclassical\footnote{For a comprehensive review of semiclassical techniques for multiparticle production, see~\cite{Khoze:2018mey}.} calculations of scattering amplitudes at the multiparticle threshold, in particular the $n$-particle decay of a highly virtual state $1 \to n$, suggest a rapid growth for large $n$~\cite{Cornwall:1990hh,Goldberg:1990qk,Brown:1992ay,Voloshin:1992mz,Argyres:1992np,Voloshin:1992nu,Smith:1992rq,Libanov:1994ug,Khoze:2014zha,Son:1995wz,Khoze:2017ifq}, in line with recent numerical studies~\cite{Demidov:2018czx,Demidov:2021rjp}.
As this is, naively, in contradiction to the unitarity requirement of any consistent quantum theory, these results hint at the termination of the perturbative regime or perhaps the need for novel degrees of freedom at high energies.\footnote{Phenomenologically, applied to the Higgs sector of the Standard Model, this even provides for an estimate of the energy scale of new physics~\cite{Voloshin:1992rr,Jaeckel:2014lya}.}
A prominent example of the latter is the ``Higgsplosion" mechanism advocated in~\cite{Khoze:2017tjt} (see also~\cite{Khoze:2017ifq}),
which removes physical contributions of highly off-shell states thanks to their exponentially growing $n$-particle decay rates.
This, in turn, leads to a non-trivial modification of the ultraviolet QFT dynamics~\cite{Khoze:2017lft,Khoze:2017uga,Khoze:2018bwa,Belyaev:2018mtd,Monin:2018cbi,Khoze:2018qhz}.

In multiparticle scattering processes, one may naively expect that adding higher-order vertices, $\varphi^{p}$ (with large integer $p$), to a simple renormalizable model, would drastically enhance the production rate of particles.
After all, more and more quanta will be produced increasingly rapidly from each such multiparticle vertex.
On the other hand, there is another key factor contributing to the large-$n$ growth, which is the sheer number of Feynman diagrams and the effects of their interference, as pointed out in~\cite{Khoze:2017lft}.
These effectively prefer Feynman diagrams with lower-point vertices, such as $\varphi^3$ or $\varphi^4$.
It is thus not \emph{a priori} obvious if the presence of the higher-point interactions should dramatically alter the large-$n$ behavior of amplitudes and decay rates of the original theory.
This is what we address in this work.

The higher-point vertices naturally appear in the Effective Field Theory (EFT) approach, which we will pursue.
In order to examine the high-energy behavior of multiparticle threshold amplitudes in this setting, we derive an all-orders perturbative expansion of the leading quantum corrections to the amplitude.
This series will be resummed in closed form for large $n$.
We will find that, in the EFT, multiparticle production at high energies is governed by a non-trivial combination of exponential functions.
In particular, in a model with $g \varphi^4$ and $(\lambda / \Lambda^2) \varphi^6$ interactions, the resummed $n$-point amplitudes on mass threshold ($\sqrt{s}=nm$) are of the form
\begin{equation}
	\mathcal{A}(n) \simeq \mathcal{A}_{\mathrm{tree}}(n) \, \me^{Agn^2} \cosh \left( B \frac{m}{\Lambda} \sqrt{\lambda} n^2 \right) \, .
	\label{eq:intro}
\end{equation}
We will also explain how this result is extended to the generic EFT case.

The expression~\eqref{eq:intro} is an immediate generalization of the purely quartic~\cite{Libanov:1994ug,Libanov:1995gh} and the purely sextic cases~\cite{Schenk:2021yea}.\footnote{Both results have also been established more generally in a quantum mechanical analogue, corresponding to transition amplitudes in the anharmonic oscillator~\cite{Jaeckel:2018ipq,Jaeckel:2018tdj,Schenk:2019kmx}.}
Here, the sign of the exponent $A$ is crucial.
If it is positive, the multiparticle threshold amplitudes grow exponentially for large $n$.
In contrast, if it is negative, they may be exponentially suppressed, if the quartic interaction is parametrically large compared to the sextic self-coupling.
That is, phenomenologically, their intricate interplay will lead to either an exponential suppression or exponential enhancement of the amplitude in the large-$n$ regime, depending on the relative sizes of all coupling constants considered in the EFT.
This provides a novel perspective on the perturbative description of scattering processes in scalar EFTs with respect to their self-interactions or even their high-energy cut-off scale.

It is important to note that there are perturbative corrections that are not captured by the large-$n$ resummation of loop diagrams, that we employ to derive~\eqref{eq:intro}.
These higher-order corrections are governed by higher powers of the effective expansion parameters $gn$ and $(\sqrt{\lambda}n)^2$.
In the large-$n$ regime where these parameters are small, these corrections are negligible and~\eqref{eq:intro} holds.
At the same time, $gn^2$ and $\sqrt{\lambda}n^2$ are large, thus justifying the presence of the exponential and hyperbolic factors in~\eqref{eq:intro}.
There is also a fully non-perturbative regime at asymptotically large $n$, where $gn$ and $(\sqrt{\lambda}n)^2$ are no longer small and one can argue that multiparticle production at high energies should be treated as intrinsically non-perturbative~\cite{Gorsky:1993ix,Son:1995wz,Khoze:2017ifq,Khoze:2018kkz,Khoze:2018mey,Dine:2020ybn}.
However, this is beyond the scope of this paper where we concentrate on the regime where the perturbative expression~\eqref{eq:intro} is valid.

This work is structured as follows.
In Section~\ref{sec:thresholdamplitudes}, we start by reviewing the generating-field technique in order to examine multiparticle production on mass threshold in scalar QFTs to arbitrary orders in perturbation theory.
We apply these methods to a simple scalar EFT model in Section~\ref{sec:eft}.
In particular, we derive a consistency condition on the field's self-interactions with respect to the perturbative approach within the EFT.
Next we go beyond the kinematic threshold and account for the effects of final state momenta in the $n$-particle phase space to estimate the corresponding multiparticle rates in the non-relativistic limit.
In Section~\ref{sec:uvcompletion}, we construct a simple ultraviolet (UV) completion of our EFT model setup and justify the validity of our results.
Finally, we briefly summarize our findings and conclude in Section~\ref{sec:conclusions}.

\section{Threshold Amplitudes from Leading Singularities}
\label{sec:thresholdamplitudes}

There are various ways to study multiparticle amplitudes associated to the production of a large number of quanta from only a few initial states.
For the purpose of our work, perhaps the most suitable technique is to make use of a generating functional.
As was shown in~\cite{Brown:1992ay}, in the case of multiparticle amplitudes at the kinematic threshold, the generating functional of tree-level amplitudes is given by a classical field solution.
In addition to being mathematically elegant, this method allows us to go beyond a tree-level computation and systematically study loop corrections to the process\footnote{One can also consider couplings to other, potentially heavy, fields~\cite{Voloshin:2017flq}.} (see, e.g.,~\cite{Voloshin:1992nu}).
Let us briefly review this formalism and consider a quantum theory of a real scalar field,
\begin{equation}
	S = \int \md^{d} x \, \left( \frac{1}{2} \left(\partial_{\mu} \varphi \right)^2 - V(\varphi) \right) \, ,
\end{equation}
where $V \left(\varphi\right)$ is some bounded potential of polynomial form, containing at least a mass term, $m^2 \varphi^2$.
We will discuss specific examples later, but let us keep the potential as general as possible for the moment.
Similarly, we will work in an arbitrary number of spacetime dimensions.

In this QFT, we aim to study multiparticle production processes of the form $1 \to n$ at the kinematic threshold.
Physically, this can be understood as the decay of a highly virtual state into $n$ particles at rest.
Indeed, the amplitude associated to this process is generated by the matrix element $\braket{0|\varphi|0}$ in the presence of a source term~\cite{Brown:1992ay},
\begin{equation}
	\mathcal{A}(n) = \left. \frac{\partial^n}{\partial z^n} \braket{0|\varphi|0} \right\rvert_{z=0} \, .
\label{eq:AnGenerate}
\end{equation}
Here, the auxiliary parameter $z$ is remnant from the field's response to a source, $J \varphi$, which can be defined as follows.
At the kinematic threshold, the source can be expanded into plane waves of a given frequency, $J(t) = J_0 \exp(i\omega t)$.
The auxiliary parameter is then given by $z_{\omega}(t) = J(t) / (m^2 - \omega^2)$.
Crucially, we can now take the simultaneous limit of the on-shell regime of the final state particles, $\omega \to m$, with vanishing source, $J_0 \to 0$, such that the auxiliary parameter remains finite, $z_{\omega} \to z(t) = z_0 \exp(i m t)$~\cite{Brown:1992ay}.
Derivatives with respect to the latter will ultimately generate the multiparticle threshold amplitudes.
For more details, we refer the reader to, e.g.,~\cite{Brown:1992ay,Khoze:2014zha}.

The expression for the multiparticle amplitude encapsulates the power of the generating-field technique.
In principle, the matrix element $\braket{0|\varphi|0}$ can be obtained systematically to arbitrary order in perturbation theory~\cite{Voloshin:1992nu} by evaluating tadpole Feynman diagrams in the background of the classical solution.
Clearly, the tree-level contribution is given by the classical field solution to the equations of motion,
\begin{equation}
	\partial_{\mu}^2 \varphi_0 + V^{\prime} (\varphi_0) = 0 \, .
\end{equation}
Here, importantly, in order for the classical field $\varphi_0$ to generate the multiparticle threshold tree-graphs according to~\eqref{eq:AnGenerate}, it has to approach the oscillatory solution $z(t) = z_0 \exp(i m t)$ when all self-interactions of the field are absent~\cite{Brown:1992ay}.
The latter also reflects the requirement that, at the kinematic threshold where all final-state particles are at rest, the classical field is spatially homogeneous.
In turn, this drastically simplifies the computation of the vacuum expectation value of the field operator in the presence of a source at any order in perturbation theory.
In fact, the perturbative expansion of the multiparticle amplitude can be systematically assessed and features an intriguing structure in the large-$n$ regime, as we will discuss in the following.

\subsection{Leading quantum corrections to all orders in perturbation theory}

In order to examine the inherent structure of the generating matrix element~\eqref{eq:AnGenerate} (and similarly for the associated multiparticle amplitude) to all orders in perturbation theory, let us closely follow the formalism outlined in~\cite{Libanov:1996vq}.
For simplicity, we normalize all quantities to the mass of the scalar particles, by setting $m = 1$.

In general, one can start by arguing that the classical field will exhibit a singularity at a certain time $t_s$.
Naively, this is because the associated $1 \to n$ process, strictly speaking, does not conserve energy.
The initial state has to be highly virtual and therefore of sufficient energy in order to produce a large number of quanta from the vacuum.
Formally, this energy is provided by the insertion of the field operator at some spacetime point, as the amplitude corresponds to the object $\braket{n|\varphi|0}$ (see, e.g.,~\cite{Libanov:1994ug,Son:1995wz}).
In turn, the singularity structure of the field is responsible for the well-known factorial growth of multiparticle amplitudes.
This can, for instance, be seen as follows.
We can consider the generating matrix element as a function of the source, which we expect to be endowed with a perturbative expansion in the complex $z$-plane.
In this scenario, the Cauchy-Abel theorem relates the finite radius of convergence of this expansion to the large-order asymptotics of its coefficients.
In particular, the large-$n$ behavior of the multiparticle amplitude~\eqref{eq:AnGenerate} is determined by the position of the singularity $z_s$ nearest to $z=0$ in the complex $z$-plane of the vacuum expectation value of the field operator.
More precisely, this means that, for large $n$, the amplitude is of the form~\cite{Goldberg:1990ys,Voloshin:1992mz}
\begin{equation}
	\abs{\mathcal{A}(n)} \sim n! \frac{1}{\abs{z_s}^n} \, .
\label{eq:cauchyabel}
\end{equation}
As the source itself is a function of time, this implies that the asymptotic behavior of the amplitude at large $n$ is determined by its leading singularity in the time variable, $t_s$.
Hence, the singularity structure, induces the growth of $\mathcal{A}(n)$ at large orders in perturbation theory.
This observation will be key in our following argument.

In the analysis of the leading singularity, it is useful to Wick-rotate the field's domain to real (Euclidean) time,
\begin{equation}
	\tau = it + \ln z_0 + c \, ,
\end{equation}
where, in general, $c$ involves all coupling constants of the theory.
In addition, because of time-translation invariance, $c$ can always be chosen such that the field is singular at the origin, $\tau_s = 0$.
Consequently, in the vicinity of this singularity, the classical field can be expanded as
\begin{equation}
	\varphi_0 \left(\tau\right) \simeq \varphi_s\left(\tau\right) + \ldots \, .
\end{equation}
Here, $\varphi_s$ denotes the most singular term of the expansion as $\tau \to \tau_s$, while the dots represent less singular terms.
In fact, as we will demonstrate next, the leading singularity of the classical field will allow us to determine \emph{all} leading quantum corrections to the generating matrix element.

\bigskip

In general, in order to obtain the quantum corrections to the generating function of the multiparticle threshold amplitudes, we can decompose the field into a classical background and its quantum fluctuations, $\varphi = \varphi_0 + \phiq$.
Similarly, the matrix element takes the form
\begin{equation}
	\braket{0|\varphi|0} = \varphi_0 + \braket{0|\phiq|0} \, .
\label{eq:vev_decomp}
\end{equation}
The quantum fluctuations can then be evaluated perturbatively via tadpole diagrams of the quantum field $\phiq$ propagating in the classical background $\varphi_0$.
For instance, the first quantum correction to the vacuum expectation value is illustrated in Fig.~\ref{fig:1-loop}.
In practice, for a computational evaluation, we have to know the propagator associated to $\phiq$, given by
\begin{equation}
	\left(\partial_{\mu}^2 + V^{\prime\prime} (\varphi_0) \right) D (x, y) = \delta^d (x-y) \, .
\label{eq:propagator}
\end{equation}
In order to obtain $D(x,y)$, in principle, we need to invert the differential operator on the left hand side of this equation.
Fortunately, the background field is spatially homogeneous, thereby drastically simplifying this operation.
For instance, it is useful to consider the mixed time-momentum representation of the propagator~\cite{Voloshin:1992nu,Libanov:1994ug},
\begin{equation}
	D \left(x, x^{\prime}\right) = \int \frac{\md^{d-1} \vec{p}}{\left(2\pi\right)^{d-1}} \, \me^{i \vec{p} \left(\vec{x} - \vec{x}^{\prime}\right)} D_{\vec{p}} \left(t,t^{\prime}\right) \, .
\end{equation}
Using this representation and performing the Wick rotation to Euclidean time, the propagator equation can be written as
\begin{equation}
	\left(- \partial_\tau^2 + \vec{p}^2 + V^{\prime\prime} (\varphi_0) \right) D_{\vec{p}} \left(\tau, \tau^{\prime}\right) = \delta\left(\tau - \tau^{\prime}\right) \, ,
\end{equation}
which, as a standard result of ordinary differential equations, is solved by
\begin{equation}
	D_{\vec{p}} \left(\tau, \tau^{\prime}\right) = \frac{1}{W_{\vec{p}}} \left[f_1^{\omega}\left(\tau\right) f_2^{\omega}\left(\tau^{\prime}\right) \theta\left(\tau^{\prime} - \tau\right) + f_1^{\omega}\left(\tau^{\prime}\right) f_2^{\omega}\left(\tau\right) \theta\left(\tau- \tau^{\prime}\right) \right] \, .
\end{equation}
Here, the functions $f_1^{\omega}$ and $f_2^{\omega}$ are solutions to the homogeneous equation and $W_{\vec{p}}$ is their Wronskian.
Furthermore, $\omega$ denotes the energy of each mode, $\omega^2 = \vec{p}^2 + 1$.
Unfortunately, an analytic solution of the homogeneous equation does not exist for an arbitrary potential term.
Nevertheless, we can still determine the properties of the propagator in different regimes of $\tau$.
For instance, in the vicinity of the singularity (the most important regime for the remaining part of this work), we first note that $V^{\prime\prime} \left(\varphi_0\right) = \varphi_s^{\prime\prime\prime} / \varphi_s^{\prime}$ as we approach the singular point, $\tau \to \tau_s = 0$~\cite{Libanov:1996vq}.
This is an immediate consequence of the equations of motion of the field.
Up to some normalization, this implies that in terms of the leading singularity, the homogeneous solution in the vicinity of $\tau_s$ can be written as
\begin{equation}
	f \left(\tau\right) \simeq \varphi_s^{\prime} \left(\tau\right) + \ldots \, ,
\end{equation}
where the dots represent less singular terms.
Therefore, the propagator in the mixed time-momentum representation schematically reads~\cite{Libanov:1996vq}
\begin{equation}
	D_{\vec{p}} \left(\tau, \tau^{\prime}\right) \simeq \frac{1}{W_{\vec{p}}} \varphi_s^{\prime} \left(\tau\right) \varphi_s^{\prime} \left(\tau^{\prime}\right) \, .
\label{eq:propagatorsingularity}
\end{equation}
We again remark that this equality only holds in the vicinity of the leading singularity of the background field.
Further note that, as the homogeneous propagator equation does not contain a friction term, $\partial_\tau f(\tau)$, the Wronskian does not depend on time.
It solely captures information on the couplings as well as the momenta of the process.

\bigskip

This simple expression for the propagator, in terms of the leading singularity of the background field, is what enables us to obtain all leading quantum corrections to the generating matrix element~\eqref{eq:AnGenerate}.
In particular, as we will point out momentarily, the leading singularity of the field is in one-to-one correspondence with the leading-$n$ quantum corrections to the amplitude at multiparticle threshold.
Therefore, if we are interested in the large-$n$ behavior of the latter, we only have to consider the field equations in the vicinity of the leading singularity.
This property can be used to simplify the calculation of quantum corrections tremendously, because we can now consider an analogue of the quantum effective action as follows.
At each quantum-loop order, any tadpole diagram can be converted into a tree graph through appropriately cutting all internal propagators~\cite{Libanov:1994ug} (see also~\cite{Libanov:1996vq,Schenk:2021yea}).
Any additional ``external leg" obtained through this procedure can then be assigned half a power of the propagator and evaluated at tree-level, while carefully accounting for the different symmetry factors.
Importantly, the momentum integrals, present in the expression for each quantum loop, factorize, i.e.~at the $k$-th loop level the amplitude will be proportional to $\mathcal{B}^k$, where
\begin{equation}
	\mathcal{B} \propto \int \frac{\md^{d-1}\vec{p}}{\left(2\pi\right)^{d-1}} \frac{b \left(\omega\right)}{W_{\vec{p}}} \, .
\end{equation}
Here, $b \left(\omega\right)$ is some arbitrary function of the energy of each mode.
Importantly, the coefficient $\mathcal{B}$ does depend on the couplings and momenta involved in the process, but not on the quantum number $n$.
We will make use of this property later and explicitly unravel the coupling constants and the loop-momenta, e.g.~$\mathcal{B} \sim B \sqrt{\lambda}$ in a $\lambda \varphi^6$ theory.

At the same time, cutting all internal lines of the diagrams and evaluating at tree level is equivalent to solving the equations of motion for a condensate which is shifted by the cut propagator,
\begin{equation}
	-\partial_{\tau}^{2} \varphi_{\mathrm{cl}} + V^{\prime} (\varphi_{\mathrm{cl}}) = 0 \, .
\end{equation}
Here, the classical field is appropriately shifted in the vicinity of the leading singularity,
\begin{equation}
	\varphi_{\mathrm{cl}} (\tau) = \varphi_s (\tau) + \sqrt{\mathcal{B}} \varphi_s^{\prime} (\tau) + \ldots \, .
\end{equation}
That is, finding the solution to this boundary value problem is equivalent to summing tree graphs in a theory where all internal propagators are appropriately cut.
Indeed, the desired solution is simply given by a Taylor expansion of the leading singularity around the condensate shift~\cite{Libanov:1996vq},
\begin{equation}
	\varphi_{\mathrm{cl}} (\tau) = \varphi_s \left(\tau + \sqrt{\mathcal{B}}\right) = \sum_{k=0}^{\infty} \frac{\sqrt{\mathcal{B}}^{k}}{k!} \frac{\partial^k \varphi_s}{\partial\tau^k} \, .
\end{equation}
Identifying both procedures and matching their coefficients (while carefully counting all tree-level graphs and their associated symmetry factors) will lead to an additional factor of $(2k)!/(2^kk!)$ in the perturbative coefficients of the generating function (for details see, e.g.,~\cite{Libanov:1994ug,Libanov:1996vq}).
Therefore, we finally arrive at the series expansion for the generating matrix element~\cite{Libanov:1996vq}
\begin{equation}
	\braket{0|\varphi|0} = \sum_{k=0}^{\infty} \frac{1}{k!} \left(\frac{\mathcal{B}}{2}\right)^k \frac{\partial^{2k} \varphi_s}{\partial \tau^{2k}} \, .
\label{eq:MatrixElementSeries}
\end{equation}
We again remark that this has to be understood as an expansion in terms of the leading singularity of the classical background field.
Remarkably, the latter has enabled us to write down full perturbative expansion of the leading quantum corrections to generating matrix element.
This series can then be resummed in order to obtain the leading-$n$ contributions to the multiparticle threshold amplitudes to all orders in perturbation theory, as we will discuss next.

\subsection{Exponentiation of the multiparticle amplitude}

The previous result demonstrates the power of the generating-field technique with respect to studying multiparticle production beyond tree level.
At this point, in order to take full advantage of the expression for the generating matrix element~\eqref{eq:MatrixElementSeries}, we aim to verify our earlier claims that the leading singularity determines the leading-$n$ contributions to the multiparticle threshold amplitude.

While the Cauchy-Abel theorem~\eqref{eq:cauchyabel} already gives an abstract argument why this is the case, let us nevertheless add a somewhat more explicit justification, also put forward in~\cite{Libanov:1996vq}.
We first note that, by Cauchy's differentiation formula, we can write for the amplitude
\begin{equation}
	\mathcal{A}(n) = \left. \frac{\partial^n}{\partial z^n} \braket{0|\varphi|0} \right\rvert_{z=0} = \frac{n!}{2\pi i} \oint \frac{\md z}{z^{n+1}} \braket{0|\varphi|0} \, ,
\label{eq:AnCauchy}
\end{equation}
where the integrand is given by the perturbative expansion~\eqref{eq:MatrixElementSeries}.
Let us now assume that the leading singularity of the background field is schematically of the form $\varphi_s (\tau) \sim 1 / \tau^{\alpha}$ for some positive exponent, $\alpha > 0$.
This assumption is indeed justified, since we can always shift the singularity of the field to the origin due to time-translation invariance.
Under this assumption, the derivative terms of the perturbative expansion of the generating matrix element then read $\partial_\tau^{2k} \varphi_s \sim 1/\tau^{2k+\alpha}$.
These can be evaluated in the Cauchy contour integral~\eqref{eq:AnCauchy}, yielding
\begin{equation}
	\frac{n!}{2\pi i} \oint \frac{\md z}{z^{n+1}} \frac{\partial^{2k} \varphi_s}{\partial \tau^{2k}} \sim n! \int_c^{2\pi i + c} \md \tau \, \frac{\me^{-\tau n}}{\tau^{2k+\alpha}} \sim n! \, n^{2k+\alpha-1} \Gamma(1-2k-\alpha) \, .
\end{equation}
That is, naively, each derivative of the leading singularity will add a power of $n$.
In addition, we observe that it is indeed the \emph{leading} singularity, which in our example is parametrized by the positive constant $\alpha$, that gives the leading quantum contribution to the multiparticle threshold amplitude in the large-$n$ regime.\footnote{Any less singular term, i.e.~terms of powers smaller than $\alpha$, would contribute terms proportional to powers of $n$ that are similarly smaller than $\alpha$.}
Most importantly, by evaluating the full perturbative expression within the contour integral~\eqref{eq:AnCauchy}, this observation finally implies that the multiparticle threshold amplitude is of exponential form~\cite{Libanov:1996vq},
\begin{equation}
	\mathcal{A}(n) \simeq \mathcal{A}_{\mathrm{tree}}(n) \exp \left( \mathcal{B} n^2 \right) \, ,
\label{eq:AnExponential}
\end{equation}
where $n$ is large, $n \gg 1$.
We again remark that the coefficient $\mathcal{B}$ depends on the number of spacetime dimensions, through the loop momentum integrals, and hence contains information on the renormalization properties of the theory.
Furthermore, it is a function of all coupling constants.
We also note that, physically, we indeed expect an additional factor of $n^2$ for each quantum correction of a given multiparticle process with $n$ particles in the final state.
Pictorially speaking, for a diagram with $n$ external legs in the final state, we have approximately $n^2$ possibilities to form a closed loop out of the external quanta.

\bigskip

In summary, if we isolate the leading singularity from the classical background field, we can compute (and resum) the leading-$n$ contribution to the multiparticle threshold amplitude to all orders in perturbation theory.
The result is a remarkably simple exponential expression.
While this seems impressive in its own right, there is still some peculiar subtlety of the generating-field technique as applied above.
That is, the exponential function~\eqref{eq:AnExponential} is only formally correct in the large-$n$ limit, where a distinction between different integer values of $n$ is lost.
However, in practice, a distinction between, say, $n = 999$ and $n = 1001$ is important, as we will point out momentarily.
Without this distinction, the exponential resummation of the perturbative expansion may induce an overcounting of quantum corrections at certain loop orders.

\subsection{Overcounting quantum corrections}

According to our previous discussion, multiparticle amplitudes at the kinematic threshold have to exponentiate at high energies.
We will now argue that a simple exponentiation of these amplitudes is too naive in order to represent their correct large-$n$ behavior.
In fact, the exponentiation has to be replaced by a more complicated combination of functions, all of which are generated by \emph{separate} matrix elements.
Evidence for this has already emerged in~\cite{Schenk:2021yea}, where different complex branches of the classical background field were identified that cancel quantum contributions at certain loop orders in order to account for the difference in their generating functions.

In order to illustrate the problem, let us consider a toy theory of a real scalar field with a sextic self-interaction, as has been used in~\cite{Schenk:2021yea},
\begin{equation}
	S = \int \md^{d} x \, \left( \frac{1}{2} \left(\partial_{\mu} \varphi \right)^2 - \frac{m^2}{2} \varphi^2 - \frac{1}{6} \frac{\lambda}{\Lambda^{2(d-3)}} \varphi^6 \right) \, .
\label{eq:ActionPhi6}
\end{equation}
Here, we have introduced the UV cut-off scale $\Lambda$ in order to make the coupling constant $\lambda$ dimensionless.
For simplicity, let us consider a positive mass parameter, $m^2>0$, corresponding to an unbroken $\mathbb{Z}_2$ symmetry.
In addition, we again consider an arbitrary number of spacetime dimensions, as we are not concerned with the renormalization properties of this toy theory.
According to the result~\eqref{eq:AnExponential}, the multiparticle amplitude will be of exponential form in the large-$n$ regime.
An explicit computation verifies that this is indeed the case~\cite{Schenk:2021yea},
\begin{equation}
	\mathcal{A}(n) \simeq \mathcal{A}_{\mathrm{tree}}(n) \exp \left( B \left(\frac{m}{\Lambda}\right)^{d-3} \sqrt{\lambda} n^2 \right) \, .
\end{equation}
Here, the (dimensionless) complex coefficient $B$ depends on the number of spacetime dimensions (see also Appendix~\ref{app:coefficients} for more details).
We again stress that, in this example, we can identify the exponential coefficient of~\eqref{eq:AnExponential} as
\begin{equation}
	\mathcal{B} = B \left(\frac{m}{\Lambda}\right)^{d-3} \sqrt{\lambda} \, .
\end{equation}
Note that $\mathcal{B}$ scales with a fractional power of the coupling constant.
This is because the perturbation theory is applied in the background of the classical field solution, which, itself, comes with a non-trivial power of the coupling, $\varphi_0 \sim 1/\sqrt{\lambda}$ (for details, see~\cite{Schenk:2021yea}).
For simplicity, let us set $m = \Lambda = 1$ for the moment, as both can easily be recovered in the final result.
As was already pointed out in~\cite{Schenk:2021yea}, the above exponential expression turns out to be too naive.
This can be seen as follows.
Let us imagine to go backwards in our argument and reproduce the perturbative expansion --- order by order in terms of Feynman diagrams --- from this exponential function.
We quickly realize that all odd terms of this expansion correspond to a quantum correction with a fractional power of the coupling constant, $\sqrt{\lambda}$.
These fractional powers of the expansion parameter, however, we do not expect to be part of a perturbation theory around the global vacuum of the potential.
Instead, the reason for their appearance is that, in such naive exponentiation, different kinds of processes are confounded, as follows.

\bigskip

In theories with higher-order self-interactions all multiparticle processes have to be classified according to their number of external legs, $n$.
For instance, in the $\varphi^6$ theory~\eqref{eq:ActionPhi6}, there are two distinct classes of amplitudes.
Their distinguishing feature is the number of quantum-loop corrections that are required to generate them.
In our example, obviously, there are final states that can be reached entirely at tree-level, corresponding to particle numbers $n = 4k + 1$, for some positive integer $k$.
In contrast, some final states can only be created via loop-induced processes.
Since the six-point coupling produces four additional particles per interaction vertex, these final states differ from the former class by two quanta, $n = 4k + 3$.
This implies that, crucially, quantum corrections may change the class of a certain multiparticle amplitude.
For instance, the one-loop correction originating from a simple $1 \to 5$ tree-level process is obtained by closing two external legs in order to form a loop.
This, in turn, effectively corresponds to a $1 \to 3$ process.
In fact, in this scenario, an \emph{odd} number of loop corrections to a given multiparticle amplitude will change its classification, while an \emph{even} number of loop corrections will correspond to a process within the same subclass.
Consequently, a naive exponentiation confounds both classes of amplitudes by treating all quantum-loop corrections on the same footing.

Applied to the above example, we can immediately conclude that all odd terms of the perturbative expansion of the multiparticle amplitude correspond to a different subclass of processes as compared to the even terms.
Hence, these have to be treated separately.
At the same time, this implies that the processes representing each subclass have to be generated by different matrix elements.
In this example, the combined perturbative expansion of the generating matrix element for all multiparticle threshold amplitudes has been derived in~\cite{Schenk:2021yea}.
It is given by
\begin{equation}
	\braket{0|\varphi|0} = \varphi_0 \sum_{k=0}^{\infty} d_k \left( \lambda^{\frac{3}{2}} B \varphi_0^4 \right)^k \, ,
\end{equation}
where $\varphi_0$ denotes the classical background field solution of the theory and the coefficients of the expansion read $d_k = 2^k \Gamma(2k+1/2) / (\sqrt{\pi} k!)$.
Resumming this expression then yields the generating matrix element for \emph{both} subclasses of multiparticle amplitudes simultaneously,
\begin{equation}
	\braket{0|\varphi|0} = \frac{2 \varphi_0}{\sqrt{2\pi}} \sqrt{-\frac{1}{16x}} \me^{-\frac{1}{16x}} K_{1/4} \left(- \frac{1}{16x} \right) \, .
\end{equation}
Here, $K$ denotes the modified Bessel function of the second kind and we have defined the argument $x = \lambda^{3/2} B \varphi_0^4$.
According to the previous paragraph, this generating function does not distinguish between and hence confounds the two different classes of multiparticle processes.
We can disentangle them by noting that the even part of the matrix element generates processes with $n=4k+1$, while the odd part generates amplitudes with $n=4k+3$.
This corresponds to two genuinely different generating functions, each responsible for a different category of final states.
Consequently, we also obtain two different expressions for the multiparticle threshold amplitudes in the large-$n$ regime.
Restoring the dimensionful quantities, these are given by (see also~\cite{Schenk:2021yea})
\begin{equation}
	\mathcal{A}(n) \simeq \mathcal{A}_{\mathrm{tree}}(n) 
	\begin{cases}
		\cosh\left( B \left(\frac{m}{\Lambda}\right)^{d-3} \sqrt{\lambda}n^2\right) & n=5,9,13,\ldots \\
		\sinh\left( B \left(\frac{m}{\Lambda}\right)^{d-3} \sqrt{\lambda}n^2 \right) & n=3,7,11,\ldots
	\end{cases}
	\, .
\end{equation}
The multiparticle amplitudes corresponding to different subclasses are complementary to each other in the sense that they exhibit a similar hyperbolic structure, i.e.~a non-trivial linear combination of exponential functions.
Nevertheless, at first sight, both collapse into the naive exponentiation of the amplitude that we have seen earlier.
This is due to the fact that, in the limit $n \to \infty$, the distinction between different values of $n$ is lost.
In fact, hyperbolic functions also appear in a more general analysis of the quantum mechanical analogue, corresponding to transition amplitudes in the sextic anharmonic oscillator~\cite{Schenk:2019kmx}.

\bigskip

Let us close this discussion with a few additional remarks on the nature underlying the appearance of hyperbolic functions instead of a simple exponential in the above example.
In~\cite{Schenk:2021yea} it has been argued that the latter arise due to a residual symmetry with respect to the complex branches of the classical background field.
If the quantum fluctuations propagate in this background, the complex branches (and therefore the residual symmetry transformation) appear at each order in the loop expansion of the vacuum expectation value of the field operator.
Since they cannot be physical, they have to be integrated out.
For instance, in this example, there are four different complex branches of the background field, in turn leading to four different exponential functions in total.
These finally combine into a hyperbolic function, if all are taken into account consistently.
Here, we find that these combinations are indeed necessary, because the quantum corrections clearly distinguish between physically different multiparticle processes.
This gives further evidence that the complex branches of the classical background field play a crucial role in evaluating the generating matrix element.

\bigskip

In summary, we conclude that in scalar QFTs with higher-order self-interactions, a simple exponentiation formula for multiparticle threshold amplitudes does not hold.
Instead, in order to preserve the well-defined ordering of quantum corrections in a perturbative approach, it is replaced by a non-trivial linear combination of exponential functions.
This appears to be a generic feature of quantum theories involving scalar degrees of freedom.

\section{Multiparticle Production in a Scalar EFT}
\label{sec:eft}

We now aim to apply our results from the previous section to a more realistic scenario of a scalar EFT.
As a specific example, let us consider the quantum theory of a real scalar field with quartic and sextic self-interactions in four dimensions,
\begin{equation}
	S = \int \md^{4} x \, \left( \frac{1}{2} \left(\partial_{\mu} \varphi \right)^2 - \frac{m^2}{2} \varphi^2 - \frac{g}{4} \varphi^4 - \frac{1}{6} \frac{\lambda}{\Lambda^2} \varphi^6 \right) \, .
\label{eq:EFTaction}
\end{equation}
Here, for concreteness, we choose the mass parameter $m^2$ as well as all coupling constants to be positive.
Later, we will also discuss the model with a negative mass parameter.
Furthermore, we have introduced the UV scale $\Lambda$ such that the six-point coupling constant $\lambda$ is dimensionless.
We can interpret this model as an EFT of a real scalar field, including operators of up to dimension six, while being agnostic about the UV details of the quantum theory.
In this sense, $\Lambda$ is the high-scale cut-off of the effective theory, which suppresses all higher-order self-interactions of the field beyond the $\varphi^6$ coupling.
Later, we will argue that, in the limit where the EFT is valid, i.e.~at energies not exceeding the UV cut-off, the effects of all higher-order operators will indeed be negligible.

By carefully repeating the arguments presented in Section~\ref{sec:thresholdamplitudes}, we now want to derive the large-$n$ behavior of multiparticle processes of the form $1 \to n$ at the kinematic threshold in this EFT.
In order to do so, we need first to determine the generating functional of tree-level amplitudes, which is given by the classical solution with the appropriate boundary conditions (i.e.~it must depend holomorphically on the variable $z(t)$).
The equation of motion is
\begin{equation}
	\left( \partial_{\mu}^2 + m^2 \right) \varphi_0 + g \varphi_0^3 + \frac{\lambda}{\Lambda^2} \varphi_0^5 = 0 \, ,
\end{equation}
and one can verify that the desired solution is given by
\begin{equation}
	\varphi_0(t) = \frac{z(t)}{\sqrt{\left(1 - \frac{g}{8m^2} z^2(t)\right)^2 - \frac{\lambda}{12m^2 \Lambda^2} z^4(t)}} \, .
\label{eq:classicalfield}
\end{equation}
Here, the source parameter $z(t)$ is the oscillatory solution of the free theory, $z(t) = z_0 \exp(imt)$.
Clearly, the latter is approached by the field if all of its self-interactions are absent, $g = \lambda = 0$.
We also note that the classical solution consistently simplifies to well-known results for either vanishing six-point~\cite{Brown:1992ay} or quartic coupling~\cite{Schenk:2021yea}, respectively.

Using the classical field as the generating function for tree-level amplitudes, one can then verify that the tree-level multiparticle threshold amplitudes suffer from the familiar factorial growth,
\begin{equation}
	\mathcal{A}_{\mathrm{tree}}(n) = \left. \frac{\partial^n}{\partial z^n} \varphi_0 \right\rvert_{z=0} = n! \left(\frac{g}{8m^2}\right)^{\frac{n-1}{2}} {}_2F_1 \left( -\frac{n-1}{4}, -\frac{n-3}{4}, 1; \frac{16}{3} \frac{m^2 \lambda}{\Lambda^2 g^2}\right) \, .
\label{eq:antree}
\end{equation}
Here, $F$ denotes a hypergeometric function and, due to the $\mathbb{Z}_2$ symmetry of the theory, $n$ is necessarily odd.
Again, this expression drastically simplifies to the well-known expressions if any of the interactions vanishes (see, e.g.,~\cite{Brown:1992ay,Schenk:2021yea}).
Starting from the classical field, we can now proceed and investigate how quantum corrections contribute to the full amplitude at $n$-particle threshold by carrying out the perturbative loop expansion in the classical field background.
Using the prescription of Section~\ref{sec:thresholdamplitudes}, the leading-$n$ quantum effects are computed by expanding around the leading singularity of the classical background field above.

\subsection{Resumming the leading quantum corrections}

In principle, the classical field is the first term of a perturbative expansion of the vacuum expectation value of the field operator.
While the latter will, in general, be a complicated function of the source parameter $z(t)$, we can still determine the high-energy behavior of the associated $n$-particle amplitudes by some more general arguments.

First of all, we have already seen in Section~\ref{sec:thresholdamplitudes} that the quantum corrections (at any order in perturbation theory) are governed by the singularity structure of the quantum field propagating in the classical background.
More precisely, the leading singularity has already led us to the conclusion that the leading-$n$ quantum corrections to the multiparticle process exponentiate,
\begin{equation}
	\mathcal{A}(n) \simeq \mathcal{A}_{\mathrm{tree}} (n) \exp \left[ n^2 G(g, \lambda) \right] \, .
\label{eq:Gunknown}
\end{equation}
Here, the sub-exponential tree-level factor is given by~\eqref{eq:antree}.
Furthermore, $G$ is an \emph{a priori} unknown function of the coupling constants of the EFT.
Since it appears in the exponent, the high-energy behavior of the process is entirely determined by this function.
Although its precise expression is beyond our computational reach, for the purpose of our work, it is sufficient to establish the parametric form of $G$.
This can be done as follows.
As we have mentioned earlier, the generating matrix element can be systematically obtained through a quantum-loop expansion in tadpole graphs,
\begin{equation}
	\braket{0|\varphi|0} = \varphi_0 + \varphi_1 + \varphi_2 + \ldots \, ,
\end{equation}
where the indices indicate the corresponding loop order.
Importantly, all information on the function $G$ is necessarily contained in the first quantum correction $\varphi_1$, at least in the large-$n$ regime.
Indeed, this contribution would correspond to the linear term of the exponential series expansion associated to the multiparticle amplitude.
As we are only interested in the dependence on the EFT coupling constants, we will again set the dimensionful parts to unity, $m = \Lambda = 1$, and drop all irrelevant numerical factors.
The former can easily be recovered in the final result from a dimensional analysis.

\bigskip

\begin{figure}[t]
	\centering
	\includegraphics[width=0.3\textwidth]{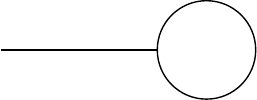}
	\caption{One-loop quantum contribution $\varphi_1$ to the vacuum expectation value of the field operator in the presence of a source, $\braket{0|\varphi|0}$.}
\label{fig:1-loop}
\end{figure}

The tadpole diagram corresponding to $\varphi_1$ is illustrated in Fig.~\ref{fig:1-loop}.
Using the Feynman rules of the quantum field propagating in the classical background $\varphi_0$, it can be schematically written as (see, e.g.,~\cite{Libanov:1994ug,Schenk:2021yea})
\begin{equation}
	\varphi_1(x) \sim \int \md x^{\prime} \, D\left(x,x^{\prime}\right) \left[ g \varphi_0\left(x^{\prime}\right) + \lambda \varphi_0^3\left(x^{\prime}\right) \right] D\left(x^{\prime},x^{\prime}\right) \, ,
\end{equation}
where $D$ denotes the propagator of the quantum field.
To extract the parametric structure of this contribution, we can first make use of the defining equation of the propagator~\eqref{eq:propagator} to obtain the differential equation
\begin{equation}
	\op \varphi_1(x) \equiv \left(\partial_{\mu}^2 + V^{\prime\prime}(\varphi_0)\right) \varphi_1(x) = \left(g \varphi_0(x) + \lambda \varphi_0^3(x)\right)D(x,x) \, .
\label{eq:diffeq_phi1}
\end{equation}
Our goal will now be to evaluate this expression in terms of the leading singularity of the background field, $\varphi_s$.
In Section~\ref{sec:thresholdamplitudes}, we have already established that the momentum modes of the propagator can be expressed as $D_{\vec{p}} \left(\tau, \tau^{\prime}\right) \simeq \varphi_s^{\prime}(\tau) \varphi_s^{\prime}\left(\tau^{\prime}\right) / W_{\vec{p}}$.
The leading singularity $\varphi_s$ can be easily extracted from the classical field~\eqref{eq:classicalfield}.
Before doing so, we first define Euclidean time through a Wick rotation and adding a constant shift,
\begin{equation}
	\tau = it \pm \ln \left( z_0 \sqrt{\frac{g}{8} \pm \sqrt{\frac{\lambda}{12}}} \right) \, .
\end{equation}
The shift moves the singularity to the origin in Euclidean time, $\tau_s = 0$.
The expansion of the classical background field around the singularity then takes the form\footnote{Note that the global complex phase in this expansion depends on the chosen sign of the constant shift in Euclidean time, after performing the Wick rotation. However, the overall phase does not play an important role in our argument.}
\begin{equation}
	\varphi_0(\tau) \simeq - \frac{i}{\sqrt{2}} \left(\frac{3}{\lambda}\right)^{\frac{1}{4}} \left(\frac{1}{\sqrt{\tau}} + \frac{\sqrt{3}}{8} \frac{g}{\sqrt{\lambda}} \sqrt{\tau} + \ldots \right) \, ,
\end{equation}
where the dots represent more regular terms of the expansion.
We can therefore identify the leading singularity as the most singular part of the above expression,
\begin{equation}
	\varphi_s(\tau) \simeq - \frac{i}{\sqrt{2}} \left(\frac{3}{\lambda}\right)^{\frac{1}{4}} \frac{1}{\sqrt{\tau}} \, .
\end{equation}
The derivative of $\varphi_0$ (and its representation in terms of the leading singularity $\varphi_s$) is the key ingredient for the parametric form of the propagator as given in~\eqref{eq:propagatorsingularity}.
Similarly, the Wronskian contributes certain powers of the coupling constants to the final propagator result.
In particular, expanding its inverse in the couplings, it schematically reads $1/W_{\vec{p}} \sim \sqrt{\lambda} + g + \mathcal{O}(g^2/\sqrt{\lambda})$.
Therefore, combining this result with the expansion in terms of the leading singularity, the propagator modes at coinciding times parametrically take the form
\begin{equation}
	D_{\vec{p}}(\tau, \tau) \sim \lambda^{\frac{3}{2}} \varphi_s^6 + g \lambda \varphi_s^6 + g \sqrt{\lambda} \varphi_s^4 + g^2 \varphi_s^4 + \ldots \, ,
\end{equation}
where the dots again represent less singular terms.
We can now plug this parametric form into the differential equation~\eqref{eq:diffeq_phi1}.
This yields
\begin{equation}
	\op \varphi_1 \sim \lambda^{\frac{5}{2}} \varphi_s^9 + g \lambda^2 \varphi_s^9 + g \lambda^{\frac{3}{2}} \varphi_s^7 + g^2 \lambda \varphi_s^7 + \ldots \, ,
\end{equation}
where we omitted numerical factors that contain information on the inherent loop momentum integrals.
From this singularity structure of the leading quantum correction, we can finally determine the first term of the perturbative expansion of the generating matrix element $\braket{0|\varphi|0}$.
That is, similar to the methods presented in~\cite{Libanov:1994ug,Schenk:2021yea}, we can invert the operator on the left hand side by noting that, in the vicinity of the leading singularity, $\op^{-1} \varphi_0^k \sim \varphi_0^{k-4} / \lambda$, such that schematically
\begin{equation}
	\varphi_1 \sim \lambda^{\frac{3}{2}} \varphi_s^5 + g \lambda \varphi_s^5 + g \sqrt{\lambda} \varphi_s^3 + g^2 \varphi_s^3 + \ldots \, .
\label{eq:phi1_parametric}
\end{equation}
In a next step, in order to obtain the one-loop quantum correction to the multiparticle amplitude, $\mathcal{A}_1$, we need to take the $n$-th derivative of $\varphi_1$ with respect to the source parameter $z(t)$.
Clearly, in the vicinity of the leading singularity, we can identify the classical background field with its most singular term, $\varphi_0 \simeq \varphi_s$.
Therefore, using~\eqref{eq:phi1_parametric}, we can equivalently consider the derivatives of a certain power of the background field $\varphi_0$.
More generally, at the $k$-th loop order, we can schematically identify the contributions to the multiparticle amplitude as $\mathcal{A}_k = \partial_z^n \varphi_k \simeq \partial_z^n \varphi_0^{n_k}$.
We note that these are given by
\begin{equation}
	 \left. \frac{\partial^n}{\partial z^n} \varphi_0^{n_k} \right\rvert_{z=0} = n! \left(\frac{g}{8m^2}\right)^{\frac{n-1}{2}} \left(\frac{8m^2}{g}\right)^{\frac{n_k-1}{2}} {}_2F_1 \left( \frac{n_k-n}{4}, \frac{n_k-n}{4} + \frac{1}{2}, \frac{n_k+1}{2}; \frac{16}{3} \frac{m^2 \lambda}{\Lambda^2 g^2}\right) \, .
\end{equation}
Here, both $n$ and $n_k$ have to be odd integers in order to give a non-vanishing result.
Naively, this is again due to the $\mathbb{Z}_2$ symmetry of the model.
We remark that the above expression is an immediate generalization of the tree-level result~\eqref{eq:antree}, beyond $n_k = 1$.
Applied to the singular expansion of $\varphi_1$, we find that the terms drastically simplify in the large-$n$ regime,
\begin{equation}
	\left. \frac{\partial^n}{\partial z^n} \varphi_0^{3} \right\rvert_{z=0} \sim \mathcal{A}_{\mathrm{tree}}(n) \frac{n}{\sqrt{\lambda}} \, , \quad \left. \frac{\partial^n}{\partial z^n} \varphi_0^{5} \right\rvert_{z=0} \sim \mathcal{A}_{\mathrm{tree}}(n) \frac{n^2}{\lambda} \, .
\end{equation}
Combining this result with the schematic form of the one-loop tadpole contribution~\eqref{eq:phi1_parametric}, we finally arrive at the parametric expression for the first quantum correction to the multiparticle amplitude at the kinematic threshold, which for large $n$ reads
\begin{equation}
	\mathcal{A}_1 = \left. \frac{\partial^n}{\partial z^n} \varphi_1 \right\rvert_{z=0} \sim \mathcal{A}_{\mathrm{tree}} (n) \left( A g n^2 + B \sqrt{\lambda} n^2 + C g n + D \frac{g^2}{\sqrt{\lambda}} n \right) \, .
\end{equation}
Here, we have introduced the complex constants $A$ to $D$.
In the large-$n$ regime, we find that the terms proportional to $C$ and $D$ are subleading as compared to the first two contributions.
For instance, this is true in the simultaneous double scaling limit, $g, \lambda \to 0$ and $n \to \infty$, while keeping the combinations $\eta = gn$ and $\kappa = \sqrt{\lambda} n$ fixed.\footnote{In this limit, the contribution can be written as $\sim A \kappa^2 / \sqrt{\lambda} + B \eta^2 / g + C \eta^2 / (g n) + D \eta^2 / \kappa$. Therefore, the terms proportional to $C$ and $D$ are either suppressed by powers of $n$ or constant, respectively.}
Therefore, we find that the exponent $n^2 G$ in~\eqref{eq:Gunknown} factorizes into separate contributions from both couplings, $g$ and $\lambda$.
In particular, in the large-$n$ regime, we can identify the exponent function as
\begin{equation}
	G(g,\lambda) \simeq Ag + B \sqrt{\lambda} \, .
\label{eq:expG}
\end{equation}
We remark that this form of the exponent is in exact agreement with previous works on the scenarios $g=0$ or $\lambda = 0$~\cite{Libanov:1994ug,Schenk:2021yea}.
The high-energy behavior of the multiparticle process is then determined by the exponential function,
\begin{equation}
	\mathcal{A}(n) \simeq \mathcal{A}_{\mathrm{tree}} (n) \exp \left[ A g n^2 + B \sqrt{\lambda} n^2 \right] \, .
\end{equation}
Although we have obtained a fully resummed, remarkably simple expression for the amplitude, we are still facing the problem that its exponential form is overcounting the quantum corrections at each loop order, as we have already discussed in Section~\ref{sec:thresholdamplitudes}.
That is, similar to the purely sextic example (where $g=0$), we again have to distinguish between two different classes of multiparticle processes, i.e.~between final states with $n=5,9,13,\ldots$ as well as final states with $n=3,7,11,\ldots \,$.
The fact that the six-point coupling enters the expansion with a fractional power indicates that the exponential function overcounts the quantum corrections associated to this interaction (see Section~\ref{sec:thresholdamplitudes} for a detailed explanation).
Therefore, depending on the class of multiparticle processes we are considering, we have to take into account solely the even or odd terms of the perturbative series in $\lambda$, respectively.
As the quartic self-interaction does not distinguish between these, the exponentiation with respect to $g$ remains valid.
Therefore, if we consider final states with particle numbers $n = 5, 9, 13, \ldots \,$, the leading-$n$ quantum corrections will resum into the expression
\begin{equation}
	\mathcal{A}(n) \simeq \mathcal{A}_{\mathrm{tree}}(n) \, \me^{Agn^2} \cosh \left( B \sqrt{\lambda} n^2 \right) \, .
\label{eq:anresummed}
\end{equation}
Similarly, if we consider states with $n = 3, 7, 11, \ldots$, the hyperbolic cosine is replaced with its sine counterpart.
Naively, this can be understood as an immediate generalization of earlier results~\cite{Libanov:1994ug,Schenk:2021yea}.
Finally, for the remaining part of this work, we will be mostly interested in the consequences of this expression.
For this, we can easily recover the mass scales $m$ and $\Lambda$ present in the original theory, effectively by rescaling $\lambda \to m^2/\Lambda^2 \lambda$, and hence we obtain
\begin{equation}
	\mathcal{A}(n) \simeq \mathcal{A}_{\mathrm{tree}}(n) \, \me^{Agn^2} \cosh \left( B \,\frac{m}{\Lambda} \sqrt{\lambda} n^2 \right) \, .
\label{eq:anresummed2}
\end{equation}
Note that, here, the complex coefficients $A$ and $B$ are dimensionless.
The above expression for the multiparticle threshold amplitude has some interesting consequences at high energies.
Let us investigate some of these in detail in the following.

\subsection{The breakdown of resummed perturbation theory in the EFT}

The leading-$n$ resummed multiparticle threshold amplitude in~\eqref{eq:anresummed2} indicates a breakdown of perturbation theory at high energies.
That is, it does not necessarily decay, but, in principle, allows for a rapid growth at large $n$.
Crucially, the emergence of exponential growth or exponential suppression depends on the exponents of the $n$-particle amplitude.

Clearly, the multiparticle threshold amplitude exhibits a rapid growth, if the coefficient of the quartic interaction $A$ has a positive real part, $\Re(A) > 0$.
In fact, there is evidence that this is realized in scalar QFTs featuring a spontaneously broken symmetry~\cite{Smith:1992rq,Khoze:2017ifq,Khoze:2018kkz}.
This growth, in addition, will then necessarily be enhanced by the hyperbolic cosine originating from the sextic self-interaction of the field, even independent of the coefficient $B$.
However, a thorough treatment of this scenario would certainly require the computation of higher-order terms of the exponent in combinations of $g$, $\lambda$ and $n$.
These corrections may change the overall sign of the exponent.
However, this possibility is not supported by a semiclassical analysis~\cite{Khoze:2017ifq,Khoze:2018kkz}.

\begin{figure}[t]
	\centering
	\includegraphics[width=0.6\textwidth]{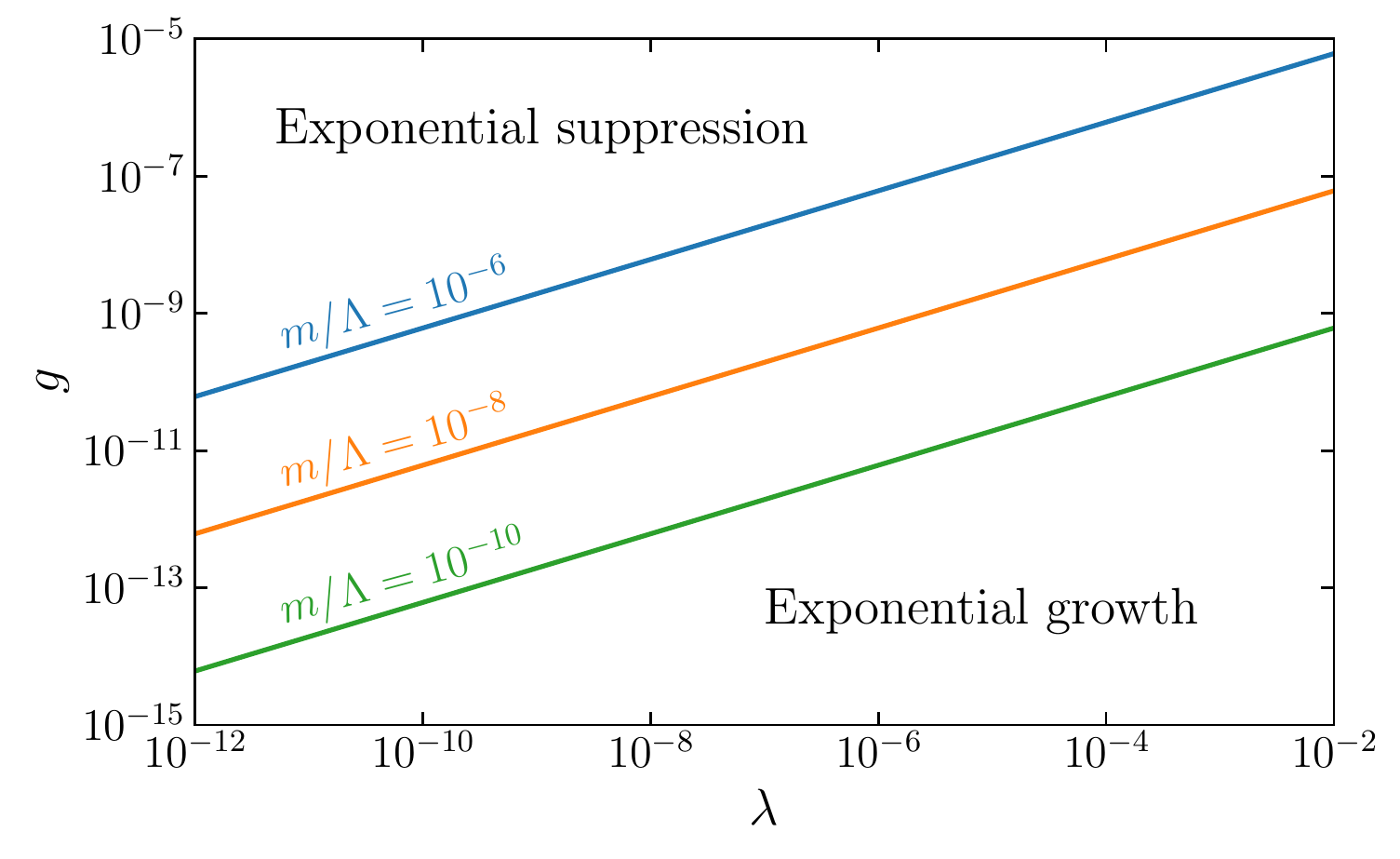}
	\caption{Parameter space of the EFT with $\Re(A) < 0$, spanned by the quartic, $g$, and sextic self-interactions, $\lambda$, of the field. Above the solid lines, the multiparticle amplitudes at the kinematic threshold are exponentially suppressed. Below the lines, they grow exponentially, thereby indicating a breakdown of (resummed) perturbation theory. The colors illustrate different ratios of energy scales of the EFT. For simplicity, here, we have neglected the tree-level contribution to the multiparticle process (see main text).}
\label{fig:higgsplosive_vs_eft}
\end{figure}

In contrast, the necessary requirement for an exponential suppression of the amplitude at high energies is that the coefficient $A$ has a negative real part, $\Re(A) < 0$.
This enables a decay of the exponential function proportional to the quartic coupling, $Ag$, for large $n$.
However, this still has to overcome the universal exponential growth originating from the hyperbolic cosine, proportional to $B \sqrt{\lambda}$.
Crucially, this growth is independent of the precise value of the coefficient $B$.
Therefore, in the large-$n$ regime, the exponential decay dominates if the coupling constants satisfy
\begin{equation}
	\text{Exponential suppression:} \quad g \gtrsim \abs{\frac{\Re(B)}{\Re(A)}} \frac{m}{\Lambda} \sqrt{\lambda} \quad \text{for} \quad \Re(A) < 0 \, .
\label{eq:ConditionExpSuppr}
\end{equation}
If this condition is satisfied, at least naively, the EFT is rendered consistent in the sense that multiparticle amplitudes at the kinematic threshold are exponentially suppressed.
Vice versa, if the condition is violated, multiparticle production processes grow without bounds at high energies.
This, in turn, would signal a breakdown of the perturbative approach.
We illustrate a few examples in Fig.~\ref{fig:higgsplosive_vs_eft}.
Here, we have explicitly used estimates for the complex coefficients $A$ and $B$, a derivation of which can be found in Appendix~\ref{app:coefficients}.
Note that, for simplicity, we have ignored the tree-level contribution to the multiparticle amplitude.
We expect this to be subdominant in the quantum number $n$ as compared to the contributions proportional to $A$ and $B$.
For instance, this is typically the case for a purely quartic or purely sextic theory.
Similarly, in the non-perturbative regime where $gn$ or $\sqrt{\lambda} n$ is large, we expect that subleading quantum corrections modify the condition~\eqref{eq:ConditionExpSuppr} accordingly.

Naively, the above condition seems to imply that we can render any EFT description consistent with respect to multiparticle production by considering an arbitrarily large scale of new physics phenomena, i.e.~by raising the UV cut-off $\Lambda \to \infty$.
Similarly, we can also argue that the quartic coupling has to be parametrically greater, by a factor $1/n$, than the six-point coupling.
To see this, we note that, in general, the EFT description is valid if the energy of the scattering process is not probing scales beyond the UV cut-off, $E \sim n m \lesssim \Lambda$.
Therefore, the particle number in the final state has to satisfy $n \lesssim \Lambda / m$.
Turning the argument around, this implies that the EFT may feature rapid growth of amplitudes at the multiparticle threshold for any relation of coupling constants of the form
\begin{equation}
	\text{Exponential growth:} \quad g \lesssim \abs{\frac{\Re(B)}{\Re(A)}} \frac{\sqrt{\lambda}}{n} \quad \text{for} \quad \Re(A) < 0 \, .
\end{equation}
Our argument guarantees that we do not explore the EFT beyond its regime of validity.
This relation seems to require a certain hierarchy of coupling constants within the EFT framework.
While this may introduce another fine-tuning problem into the realm of viable scalar field theories, we note that these hierarchies can nevertheless be engineered in very simple UV completions of the EFT.

\bigskip

Let us close this discussion with a few words of caution.
We again emphasize that there is no possibility for an exponential suppression of the amplitude at the multiparticle threshold, if the real part of the coefficient $A$ is positive (as is the case for theories with a broken symmetry in the perturbative regime).
Therefore, the sign of $A$ is crucial for the phenomenological implications of our result.
Such scenario would inevitably feature rapidly growing multiparticle threshold amplitudes, at least in perturbation theory.
If this behavior also persists beyond the kinematic threshold, this may signal a fundamental non-perturbative change of the UV dynamics of the theory~\cite{Khoze:2017tjt,Khoze:2017ifq,Khoze:2017lft}.
Thus, in the next section, we examine this scenario away from the multiparticle threshold.

\subsection{Moving away from the kinematic threshold}

In principle, an exponential growth of multiparticle production amplitudes at the kinematic threshold does not pose any catastrophic problem for the UV dynamics and ultimately the consistency of the quantum theory, because the associated cross section is zero due to its vanishing phase space volume.
Therefore, physically, a mechanism similar to ``Higgsplosion"~\cite{Khoze:2017tjt} can only be realized in a scalar QFT, if the drastic phase space suppression is also overcome by the rapid growth of the matrix element associated to the process.
These two factors are hence in competition, in order to eventually enable or disable ``Higgsplosive" behavior at high energies.

While the computation of the fully phase-space-integrated multiparticle production cross section is beyond our reach, it is possible to move away from the kinematic threshold into a non-relativistic regime (see, e.g.,~\cite{Libanov:1994ug,Khoze:2014kka,Khoze:2017ifq,Khoze:2018kkz}).
The relevant deformation parameter in this scenario is the average kinetic energy per particle and mass,
\begin{equation}
	\varepsilon = \frac{E - nm}{nm} \, .
\end{equation}
In the non-relativistic regime, $\varepsilon$ is taken to be small, $\varepsilon \ll 1$.
For instance, using this deformation, the tree-level amplitude away from the kinematic threshold gets an exponentially small correction in $\varphi^4$ theory in the unbroken phase~\cite{Libanov:1994ug},
\begin{equation}
	\mathcal{A}_{\mathrm{tree}}(n, \varepsilon) \simeq \mathcal{A}_{\mathrm{tree}}(n, \varepsilon=0) \, \me^{-\frac{5}{6} n \varepsilon} \, .
\label{eq:AnTreeOffThreshold}
\end{equation}
A similar behavior also holds in the broken phase and in presence of gauge fields~\cite{Khoze:2014kka}.
However, this contribution to the multiparticle process is subdominant compared to the effect of the tiny phase space volume as $\varepsilon \to 0$.
In a four-dimensional theory, the latter is proportional to $\varepsilon^{3n/2}$.
This can trivially be lifted into the exponent of the naive exponentiation~\eqref{eq:Gunknown}, such that~\cite{Khoze:2017ifq,Khoze:2018kkz}
\begin{equation}
	n^2 G \to n^2 G + \frac{3}{2} n \left( \log \frac{\varepsilon}{3\pi} + 1\right) + \mathcal{O} \left( \varepsilon \right) \, .
\end{equation}
Here, we have omitted terms proportional to powers of $\varepsilon$, e.g.~originating from the tree-level correction~\eqref{eq:AnTreeOffThreshold}, as the logarithmic term dominates in the non-relativistic regime, where $\varepsilon \ll 1$.
The latter term is what needs to be subdominant with respect to the first part, $n^2 G$, in order to avoid an exponential suppression from the phase space factor and hence, ultimately, enable ``Higgsplosive" behavior of the theory.
Therefore, plugging in the expression for $G$ from~\eqref{eq:expG} and using the EFT validity condition, $m/\Lambda \lesssim 1/n$, we arrive at the schematic form for the exponent
\begin{align}
	n^2 G &\simeq n^2 \left(A g + B \frac{m}{\Lambda} \sqrt{\lambda}\right) + \frac{3}{2} n \left( \log \frac{\varepsilon}{3\pi} + 1\right) \\
	&\lesssim A g n^2 + B \sqrt{\lambda} n + \frac{3}{2} n \left( \log \frac{\varepsilon}{3\pi} + 1\right) \, ,
\end{align}
which is valid for fixed $\varepsilon \ll 1$ and large $n$.
In this regime, we find that the term proportional to the six-point coupling $\lambda$ comes with the same power of $n$ as the phase space factor and can therefore never be dominant for $\epsilon \to 0$.
Consequently, any form of ``Higgsplosive" behavior can indeed only be achieved through the quartic coupling term $Ag$, which exhibits an additional power of the quantum number $n$.
This, in principle, provides some chance for overcoming the drastic phase space suppression.
Therefore, we conclude that in a non-relativistic regime away from the kinematic threshold, ``Higgsplosive" behavior of the EFT can only be triggered by the renormalizable interactions.
The six-point coupling would, in turn, probe energies beyond the cut-off scale of the EFT.
If any ``Higgsplosive" behavior is triggered, however, it is \emph{always} enhanced by the hyperbolic cosine arising from the dimension-six operator of the field.
In this sense, the high-energy behavior of multiparticle processes away from the kinematic threshold is rather robust against effects of higher-dimensional operators naturally arising in an EFT setting.

\section{A Simple UV Completion}
\label{sec:uvcompletion}

So far, we have been considering a scalar EFT, while being agnostic about its high-energy degrees of freedom.
In fact, the exponentially growing contribution to the multiparticle amplitudes from higher-dimensional EFT operators may arise from renormalizable scalar QFTs in the UV.
To illustrate this claim, let us consider a simple UV-complete theory of two massive real scalar fields with quartic self-interactions, coupled by a mixing term in four dimensions,
\begin{equation}
	S = \int \md^{4} x \, \left( \frac{1}{2} \left(\partial_{\mu} \varphi \right)^2 + \frac{1}{2} \left(\partial_{\mu} \Phi \right)^2 - \frac{m^2}{2} \varphi^2 - \frac{g}{4} \varphi^4 - \frac{\Lambda^2}{2} \Phi^2 - \frac{G}{4} \Phi^4 - \frac{\lambda_P}{2} \varphi^2 \Phi^2 \right) \, .
\label{eq:UVaction}
\end{equation}
Let us assume a hierarchy of scales, where the field $\Phi$ is much heavier than $\varphi$, i.e.~$m \ll \Lambda$.
In this sense, $\varphi$ is the low-energy degree of freedom of the EFT description presented in Section~\ref{sec:eft}.
In addition, this allows us to identify the mass parameter $\Lambda$ with the cut-off scale of the EFT.
Clearly, depending on the sign of the mass parameter, $m^2$, this model features a $\mathbb{Z}_2$ symmetry that is either preserved or spontaneously broken.
Let us focus on the theory in the unbroken phase first and comment on the broken regime afterwards.

\begin{figure}[t]
	\centering
	\includegraphics[width=0.3\textwidth]{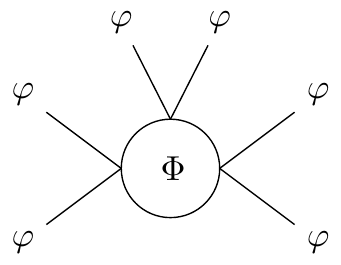}
	\hspace{3em}
	\includegraphics[width=0.25\textwidth]{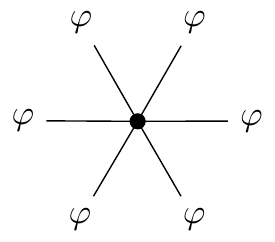}
	\caption{Coupling of six light $\varphi$-particles with a heavy $\Phi$-excitation running in the loop (left) and its corresponding six-point vertex in the EFT (right). The EFT vertex will be suppressed by the mass scale of the heavy degree of freedom.}
\label{fig:uv_eft_6pt}
\end{figure}

Due to the hierarchy of scales between the two scalar fields, we can map the UV theory to the EFT~\eqref{eq:EFTaction} by integrating out the heavy degree of freedom $\Phi$.
The portal interaction between both fields, $\lambda_P \varphi^2 \Phi^2$, will then generate all self-interactions involving arbitrary powers of $\varphi$ at low energies.
In particular, it will give rise to the sextic self-coupling of the field, as illustrated in Fig.~\ref{fig:uv_eft_6pt}.
Here, the propagator of the heavy degree of freedom running in the loop will suppress the six-point interaction by the high mass scale $\Lambda$.
Therefore, after integrating out $\Phi$, we can determine the six-point vertex as proportional to a cubic power of the portal coupling, $ (\lambda_P^3 / 16\pi^2) \varphi^6 / \Lambda^2$.
Translating this into the action of the EFT model~\eqref{eq:EFTaction}, we can identify their six-point self-couplings,
\begin{equation}
	\lambda \simeq \frac{3 \lambda_P^3}{8 \pi^2} \, .
\end{equation}
Similarly, all higher-order self-interactions will be parametrically suppressed by additional powers of the portal coupling constant $\lambda_P$ as well as the high mass scale $\Lambda$.

While this UV theory is a very simple motivation of the EFT presented in the previous section, it also allows for a hierarchy of couplings associated to the multiparticle amplitudes.
In particular, we can, to good accuracy, neglect any self-interaction terms of the light field beyond $\varphi^6$ in the EFT.
To see this, let us briefly demonstrate that all higher-dimensional operators arising in the EFT are suppressed by powers of the quantum number $n$.
The latter is considered to be large, but for the EFT description to be valid it has to remain below the UV scale.
This guarantees that the $n$-particle processes do not probe the high-energy degrees of freedom, $E \sim nm \lesssim \Lambda$.
As an example, let us compare the contributions to the process from six-point and eight-point interactions.
These emerge in the scalar potential of the EFT as $(\lambda_6 / \Lambda^2) \varphi^6$ and $(\lambda_8 / \Lambda^4) \varphi^8$, respectively.
According to the arguments outlined in Section~\ref{sec:eft}, in the naive exponentiation of multiparticle threshold amplitudes, both interactions would appear in the exponent as
\begin{equation}
	n^2 G \sim n^2 \left( \frac{m}{\Lambda} \sqrt{\lambda_6} + \left(\frac{m}{\Lambda}\right)^{\frac{4}{3}} \sqrt[3]{\lambda_8} \right) \, .
\end{equation}
Here, the third root of the eight-point coupling constant $\lambda_8$ is essentially due to the presence of three subclasses of multiparticle processes originating from an eight-point vertex.
The associated overcounting of quantum corrections is then removed by suitably combining the exponential functions~\cite{Schenk:2021yea}.
By describing the EFT couplings in terms of their UV counterpart, $\lambda_6 \sim \lambda_P^3$ and $\lambda_8 \sim \lambda_P^4$, and using the EFT validity condition, $n \lesssim \Lambda / m$, we can then rewrite the exponent as
\begin{equation}
	n^2 G \lesssim n^2 \left( \frac{1}{n} \lambda_P^{\frac{3}{2}} + \frac{1}{n^{\frac{4}{3}}} \lambda_P^{\frac{4}{3}} \right) \, .
\end{equation}
Therefore, we find that, e.g., in the double scaling limit $\lambda_P \to 0$ and $n \to \infty$ with $\lambda_P^{3/2} n$ fixed,\footnote{This is a somewhat obscured reformulation of the double scaling limit we have already considered in the previous section.} the second term corresponding to the eight-point coupling is suppressed by a factor of approximately $\sim 1/n^2$ as compared to the six-point interaction.
Similarly, all higher-dimensional operators, such as $(\lambda_{10} / \Lambda^6) \varphi^{10}$, can be discarded in this regime.
This observation therefore motivates and justifies our initial EFT model presented in Section~\ref{sec:eft}.

Similarly, the same observations hold for a theory featuring a spontaneously broken $\mathbb{Z}_2$ symmetry.
For instance, we could imagine to only couple the physical excitations around the field's vacuum expectation value, $v \neq 0$, via the quadratic portal interaction to the heavy field $\Phi$.
In this case, all of the above observations carry over and we would arrive at an EFT involving the light excitations that, due to the broken symmetry, now feature additional cubic interactions at low energies.
Nevertheless, the phenomenological implications of a spontaneously broken symmetry with respect to multiparticle production are drastically different, as mentioned before.

We conclude that, from the high-energy perspective, the interactions between the light and the heavy degrees of freedom via the portal coupling, $\lambda_P \varphi^2 \Phi^2$, have important implications for multiparticle processes in the corresponding EFT setting.
While one may naively expect that one can easily discard these interactions when considering amplitudes at the multiparticle threshold of the light scalar particles, our results demonstrate that this is not the case.
That is, the portal coupling gives rise to higher-order self-interactions of the light field which, in turn, will inevitably give an exponentially growing contribution to the threshold amplitude for large $n$.
Even within the validity bounds of the EFT, this indicates a breakdown of the perturbative approach.
As we have argued here, this is the case for QFTs both with or without a spontaneously broken symmetry.
We therefore may not be able to validate the perturbative behavior of multiparticle production without carefully examining the special role of a portal coupling between the light and heavy degrees of freedom.

Finally, we note that, for scattering processes of the form $2 \to n$ in this scenario, we do not expect any ``nullification" of threshold amplitudes as in, e.g., $\varphi^4$ theory~\cite{Voloshin:1992xb,Voloshin:1992nm,Argyres:1992un,Argyres:1993xa,Smith:1993hz}.
This would require a complete destructive interference between different on-shell diagrams with distinct parameters $g$, $\lambda$, $m$ and $\Lambda$ (see, e.g.,~\cite{Jaeckel:2014lya} and references therein).
It would thus not resolve the breakdown of resummed perturbation theory in the EFT setting.

\section{Conclusions}
\label{sec:conclusions}

Calculations of processes involving a large number of particles in weakly-coupled QFTs suffer (or enjoy) a breakdown of perturbation theory at high energies.
In particular, there is evidence that scalar field amplitudes at the multiparticle threshold rapidly grow with energy.
This strongly suggests the need for novel non-perturbative behavior, or perhaps even new degrees of freedom.

In this work, we have taken steps towards a more complete picture of this problem inherent to scalar QFTs at high energies.
In particular, we have investigated a simple EFT model of a real scalar field in four dimensions, featuring quartic and sextic self-interactions, $g \varphi^4$ and $(\lambda / \Lambda^2) \varphi^6$.
We find that the multiparticle amplitude at the kinematic threshold, associated to the $n$-particle decay of a highly virtual state $1 \to n$, is given by an intricate combination of exponential functions for large $n$,
\begin{equation}
	\mathcal{A}(n) \simeq \mathcal{A}_{\mathrm{tree}}(n) \, \me^{Agn^2} \cosh \left( B \,\frac{m}{\Lambda} \sqrt{\lambda} n^2 \right) \, .
\end{equation}
In general, this expression either allows for an exponential growth or an exponential suppression of the amplitude as $n$ tends to infinity, $n \to \infty$.
Remarkably, this solely depends on the quartic coefficient $A$, as the hyperbolic cosine inevitably corresponds to an exponentially growing contribution.
Clearly, the multiparticle threshold amplitudes exhibit a rapid exponential growth, if the coefficient $A$ has a positive real part, $\Re(A) > 0$, as indicated for scalar QFTs featuring a spontaneously broken symmetry.
In contrast, a necessary condition for an exponential suppression at high energies is that the quartic coefficient is negative, $\Re(A) < 0$.
In this scenario, we find a relation between the coupling constants of the EFT for large $n$.
If the quartic coupling is parametrically small compared to the six-point interaction,
\begin{equation}
	g \lesssim \frac{m}{\Lambda} \sqrt{\lambda} \lesssim \frac{\sqrt{\lambda}}{n} \, ,
\end{equation}
the multiparticle amplitudes at the kinematic threshold grow without bounds for large $n$, even within the regime where the EFT is valid.
This is illustrated in Fig.~\ref{fig:higgsplosive_vs_eft}.
That is, depending on the self-interactions of the field, the EFT may feature rapidly growing multiparticle threshold amplitudes at high energies, indicating a breakdown of the perturbative approach.
If this behavior also persists beyond the kinematic threshold, this will ultimately signal a transition to novel non-perturbative dynamics in the UV.

\section*{Acknowledgments}

We thank Joerg Jaeckel and Michael Spannowsky for useful discussions.
V.~V.~K.~is supported by the STFC under grant ST/P001246/1 and S.~S.~is funded by the Deutsche Forschungsgemeinschaft (DFG, German Research Foundation) -- 444759442.

\appendix

\section{Determining the Coefficients of the Exponentiation}
\label{app:coefficients}

In this appendix, we aim to derive the complex coefficients $A$ and $B$ that enter the multiparticle threshold amplitude for large $n$,
\begin{equation}
	\mathcal{A}(n) \simeq \mathcal{A}_{\mathrm{tree}}(n) \, \me^{Agn^2} \cosh \left( B \sqrt{\lambda} n^2 \right) \, .
\end{equation}
Both coefficients contain all information on the loop-momentum integrals of the theory, and therefore reflect its renormalization properties.
In our four-dimensional EFT model~\eqref{eq:EFTaction}, they take the schematic form
\begin{equation}
	A \propto \int \frac{\md^3 \vec{p}}{\left( 2\pi \right)^3} \frac{a(\omega)}{W_{\vec{p}}} \, , \quad B \propto \int \frac{\md^3 \vec{p}}{\left( 2\pi \right)^3} \frac{b(\omega)}{W_{\vec{p}}} \, ,
\end{equation}
where $a$ and $b$ are functions of the energy of each momentum mode, $\omega^2 = \vec{p}^2 + 1$.
Note that, here, we have normalized all mass and energy scales of the EFT to unity, $m = \Lambda = 1$.
The dependence on both can be easily recovered later.
As explained in the main text, at least for the coefficient $A$, it is important to distinguish between a theory with or without a spontaneously broken $\mathbb{Z}_2$ symmetry.
Let us focus on the former case first and briefly comment on a theory in the broken phase later.

As a simple approximation of both $A$ and $B$, we can make use of existing results.
We note that the above expression for the multiparticle amplitude drastically simplifies to well-known cases if either the quartic or the sextic coupling vanishes, $g = 0$ or $\lambda = 0$, respectively.
The latter case has for instance been established in~\cite{Voloshin:1992nu,Libanov:1994ug}.
Closely following these, we can formally write the coefficient $A$ as~\cite{Libanov:1994ug}
\begin{equation}
	A = \int \frac{\md^3 \vec{p}}{\left( 2\pi \right)^3} \, \frac{9}{8} \frac{1}{\omega \left(\omega^2 -1\right) \left(\omega^2 - 4\right)} = \frac{9}{16\pi^2} \int_{1}^{\infty} \md \omega \, \frac{1}{\sqrt{\omega^2 - 1} \left( \omega^2 - 4 \right)} \, .
\end{equation}
Clearly, the integrand is singular at the one-particle as well as the two-particle pole.\footnote{Intriguingly, the two-particle pole is relevant in the nullification of $2 \to n$ scattering amplitudes in $\varphi^4$ theory at the kinematic threshold~\cite{Voloshin:1992xb,Voloshin:1992nm,Argyres:1992un,Argyres:1993xa,Smith:1993hz}.}
While the one-particle singularity is integrable, we have to regularize the singularity associated to the two-particle pole.
This can be achieved by shifting the pole to the complex plane, $m^2 - i \epsilon$, corresponding to a deformation of the integration contour.
After the integration, we can take the limit $\epsilon \to 0$ in order to recover the regularized value of the integral.
We obtain
\begin{equation}
	A = \frac{9}{16\pi^2} \lim_{\epsilon \to 0} \int_{1}^{\infty} \md \omega \, \frac{1}{\sqrt{\omega^2 - 1 - i \epsilon} \left( \omega^2 - 4 - 4 i \epsilon \right)} = \frac{3\sqrt{3}}{64\pi^2} \left[ \ln \left(7 - 4\sqrt{3}\right) - i \pi \right] \, .
\end{equation}
The imaginary part of the integral depends on how one chooses to deform the integration contour.
Repeating the same procedure also in a theory featuring a spontaneously broken symmetry, we can summarize~\cite{Voloshin:1992nu,Smith:1992rq,Argyres:1993wz}
\begin{equation}
	A = \begin{cases}
		\frac{3\sqrt{3}}{64\pi^2} \left[ \ln \left(7 - 4\sqrt{3}\right) - i \pi \right] & \text{(unbroken theory)} \\
		\frac{\sqrt{3}}{8\pi} & \text{(broken theory)}
	\end{cases}
	\, ,
\end{equation}
Similarly, by the same regularization procedure we obtain for the coefficient of the six-point self-interaction~\cite{Schenk:2021yea}
\begin{equation}
	B = 24 \int \frac{\md^3 \vec{p}}{\left( 2\pi \right)^3} \frac{32i}{3\sqrt{3}\pi} \frac{1}{\omega \left(\omega^2 -1\right) \left(\omega^2 - 9\right)} = \sqrt{\frac{2}{3}} \frac{2}{3\pi^2} \, .
\end{equation}
In contrast to the previous case, here, we have regularized the singularity associated to the three-particle pole.

Finally, after reintroducing the dimensionful parameters of the theory, we obtain for the ratio of the real parts of both coefficients in a theory with an unbroken symmetry
\begin{equation}
	\abs{\frac{\Re(B)}{\Re(A)}} \sim 60 m \, .
\end{equation}
This quantity indeed plays an important role for the high-energy behavior of multiparticle processes at the kinematic threshold in the EFT model~\eqref{eq:EFTaction}.

\bibliographystyle{inspire}
\bibliography{refs}

\providecommand{\href}[2]{#2}\begingroup\raggedright\begin{thebibliography}{10}

\bibitem{Giudice:2008bi}
G.~F. Giudice, ``{Naturally Speaking: The Naturalness Criterion and Physics at
  the LHC},'' \href{http://arxiv.org/abs/0801.2562}{arXiv:0801.2562 [hep-ph]}.

\bibitem{Callaway:1988ya}
D.~J.~E. Callaway, ``{Triviality Pursuit: Can Elementary Scalar Particles
  Exist?},'' \href{http://dx.doi.org/10.1016/0370-1573(88)90008-7}{Phys. Rept.
  {\bfseries 167} (1988) 241}.

\bibitem{Khoze:2018mey}
V.~V. Khoze and J.~Reiness, ``{Review of the semiclassical formalism for
  multiparticle production at high energies},''
  \href{http://dx.doi.org/10.1016/j.physrep.2019.06.004}{Phys. Rept. {\bfseries
  822} (2019) 1--52}, \href{http://arxiv.org/abs/1810.01722}{[arXiv:1810.01722
  [hep-ph]]}.

\bibitem{Cornwall:1990hh}
J.~M. Cornwall, ``{On the High-energy Behavior of Weakly Coupled Gauge
  Theories},'' \href{http://dx.doi.org/10.1016/0370-2693(90)90850-6}{Phys.
  Lett. B {\bfseries 243} (1990) 271--278}.

\bibitem{Goldberg:1990qk}
H.~Goldberg, ``{Breakdown of perturbation theory at tree level in theories with
  scalars},'' \href{http://dx.doi.org/10.1016/0370-2693(90)90628-J}{Phys. Lett.
  B {\bfseries 246} (1990) 445--450}.

\bibitem{Brown:1992ay}
L.~S. Brown, ``{Summing tree graphs at threshold},''
  \href{http://dx.doi.org/10.1103/PhysRevD.46.R4125}{Phys. Rev. D {\bfseries
  46} (1992) R4125--R4127},
  \href{http://arxiv.org/abs/hep-ph/9209203}{[arXiv:hep-ph/9209203]}.

\bibitem{Voloshin:1992mz}
M.~B. Voloshin, ``{Multiparticle amplitudes at zero energy and momentum in
  scalar theory},'' \href{http://dx.doi.org/10.1016/0550-3213(92)90678-5}{Nucl.
  Phys. B {\bfseries 383} (1992) 233--248}.

\bibitem{Argyres:1992np}
E.~N. Argyres, R.~H.~P. Kleiss, and C.~G. Papadopoulos, ``{Amplitude estimates
  for multi - Higgs production at high-energies},''
  \href{http://dx.doi.org/10.1016/0550-3213(93)90140-K}{Nucl. Phys. B
  {\bfseries 391} (1993) 42--56}.

\bibitem{Voloshin:1992nu}
M.~B. Voloshin, ``{Summing one loop graphs at multiparticle threshold},''
  \href{http://dx.doi.org/10.1103/PhysRevD.47.R357}{Phys. Rev. D {\bfseries 47}
  (1993) R357--R361},
  \href{http://arxiv.org/abs/hep-ph/9209240}{[arXiv:hep-ph/9209240]}.

\bibitem{Smith:1992rq}
B.~H. Smith, ``{Summing one loop graphs in a theory with broken symmetry},''
  \href{http://dx.doi.org/10.1103/PhysRevD.47.3518}{Phys. Rev. D {\bfseries 47}
  (1993) 3518--3520},
  \href{http://arxiv.org/abs/hep-ph/9209287}{[arXiv:hep-ph/9209287]}.

\bibitem{Libanov:1994ug}
M.~V. Libanov, V.~A. Rubakov, D.~T. Son, and S.~V. Troitsky, ``{Exponentiation
  of multiparticle amplitudes in scalar theories},''
  \href{http://dx.doi.org/10.1103/PhysRevD.50.7553}{Phys. Rev. D {\bfseries 50}
  (1994) 7553--7569},
  \href{http://arxiv.org/abs/hep-ph/9407381}{[arXiv:hep-ph/9407381]}.

\bibitem{Khoze:2014zha}
V.~V. Khoze, ``{Multiparticle Higgs and Vector Boson Amplitudes at
  Threshold},'' \href{http://dx.doi.org/10.1007/JHEP07(2014)008}{JHEP
  {\bfseries 07} (2014) 008},
  \href{http://arxiv.org/abs/1404.4876}{[arXiv:1404.4876 [hep-ph]]}.

\bibitem{Son:1995wz}
D.~T. Son, ``{Semiclassical approach for multiparticle production in scalar
  theories},'' \href{http://dx.doi.org/10.1016/0550-3213(96)00386-0}{Nucl.
  Phys. B {\bfseries 477} (1996) 378--406},
  \href{http://arxiv.org/abs/hep-ph/9505338}{[arXiv:hep-ph/9505338]}.

\bibitem{Khoze:2017ifq}
V.~V. Khoze, ``{Multiparticle production in the large \ensuremath{\lambda}n
  limit: realising Higgsplosion in a scalar QFT},''
  \href{http://dx.doi.org/10.1007/JHEP06(2017)148}{JHEP {\bfseries 06} (2017)
  148}, \href{http://arxiv.org/abs/1705.04365}{[arXiv:1705.04365 [hep-ph]]}.

\bibitem{Demidov:2018czx}
S.~V. Demidov and B.~R. Farkhtdinov, ``{Numerical study of multiparticle
  scattering in $\lambda\phi^4$ theory},''
  \href{http://dx.doi.org/10.1007/JHEP11(2018)068}{JHEP {\bfseries 11} (2018)
  068}, \href{http://arxiv.org/abs/1806.10996}{[arXiv:1806.10996 [hep-ph]]}.

\bibitem{Demidov:2021rjp}
S.~V. Demidov, B.~R. Farkhtdinov, and D.~G. Levkov, ``{Numerical Study of
  Multiparticle Production in \ensuremath{\phi}$^{4}$ Theory: Comparison with
  Analytical Results},''
  \href{http://dx.doi.org/10.1134/S0021364021230028}{JETP Lett. {\bfseries 114}
  no.~11, (2021) 649--652},
  \href{http://arxiv.org/abs/2111.04760}{[arXiv:2111.04760 [hep-ph]]}.

\bibitem{Voloshin:1992rr}
M.~B. Voloshin, ``{Estimate of the onset of nonperturbative particle production
  at high-energy in a scalar theory},''
  \href{http://dx.doi.org/10.1016/0370-2693(92)90901-F}{Phys. Lett. B
  {\bfseries 293} (1992) 389--394}.

\bibitem{Jaeckel:2014lya}
J.~Jaeckel and V.~V. Khoze, ``{Upper limit on the scale of new physics
  phenomena from rising cross sections in high multiplicity Higgs and vector
  boson events},'' \href{http://dx.doi.org/10.1103/PhysRevD.91.093007}{Phys.
  Rev. D {\bfseries 91} no.~9, (2015) 093007},
  \href{http://arxiv.org/abs/1411.5633}{[arXiv:1411.5633 [hep-ph]]}.

\bibitem{Khoze:2017tjt}
V.~V. Khoze and M.~Spannowsky, ``{Higgsplosion: Solving the hierarchy problem
  via rapid decays of heavy states into multiple Higgs bosons},''
  \href{http://dx.doi.org/10.1016/j.nuclphysb.2017.11.002}{Nucl. Phys. B
  {\bfseries 926} (2018) 95--111},
  \href{http://arxiv.org/abs/1704.03447}{[arXiv:1704.03447 [hep-ph]]}.

\bibitem{Khoze:2017lft}
V.~V. Khoze and M.~Spannowsky, ``{Higgsploding universe},''
  \href{http://dx.doi.org/10.1103/PhysRevD.96.075042}{Phys. Rev. D {\bfseries
  96} no.~7, (2017) 075042},
  \href{http://arxiv.org/abs/1707.01531}{[arXiv:1707.01531 [hep-ph]]}.

\bibitem{Khoze:2017uga}
V.~V. Khoze, J.~Reiness, M.~Spannowsky, and P.~Waite, ``{Precision measurements
  for the Higgsploding Standard Model},''
  \href{http://dx.doi.org/10.1088/1361-6471/ab1a70}{J. Phys. G {\bfseries 46}
  no.~6, (2019) 065004},
  \href{http://arxiv.org/abs/1709.08655}{[arXiv:1709.08655 [hep-ph]]}.

\bibitem{Khoze:2018bwa}
V.~V. Khoze, J.~Reiness, J.~Scholtz, and M.~Spannowsky, ``{A Higgsploding
  Theory of Dark Matter},''
  \href{http://arxiv.org/abs/1803.05441}{arXiv:1803.05441 [hep-ph]}.

\bibitem{Belyaev:2018mtd}
A.~Belyaev, F.~Bezrukov, C.~Shepherd, and D.~Ross, ``{Problems with
  Higgsplosion},'' \href{http://dx.doi.org/10.1103/PhysRevD.98.113001}{Phys.
  Rev. D {\bfseries 98} no.~11, (2018) 113001},
  \href{http://arxiv.org/abs/1808.05641}{[arXiv:1808.05641 [hep-ph]]}.

\bibitem{Monin:2018cbi}
A.~Monin, ``{Inconsistencies of higgsplosion},''
  \href{http://arxiv.org/abs/1808.05810}{arXiv:1808.05810 [hep-th]}.

\bibitem{Khoze:2018qhz}
V.~V. Khoze and M.~Spannowsky, ``{Consistency of Higgsplosion in Localizable
  QFT},'' \href{http://dx.doi.org/10.1016/j.physletb.2019.01.052}{Phys. Lett. B
  {\bfseries 790} (2019) 466--474},
  \href{http://arxiv.org/abs/1809.11141}{[arXiv:1809.11141 [hep-ph]]}.

\bibitem{Libanov:1995gh}
M.~V. Libanov, D.~T. Son, and S.~V. Troitsky, ``{Exponentiation of
  multiparticle amplitudes in scalar theories. 2. Universality of the
  exponent},'' \href{http://dx.doi.org/10.1103/PhysRevD.52.3679}{Phys. Rev. D
  {\bfseries 52} (1995) 3679--3687},
  \href{http://arxiv.org/abs/hep-ph/9503412}{[arXiv:hep-ph/9503412]}.

\bibitem{Schenk:2021yea}
S.~Schenk, ``{The breakdown of resummed perturbation theory at high
  energies},'' \href{http://dx.doi.org/10.1007/JHEP03(2022)100}{JHEP {\bfseries
  03} (2022) 100}, \href{http://arxiv.org/abs/2109.00549}{[arXiv:2109.00549
  [hep-ph]]}.

\bibitem{Jaeckel:2018ipq}
J.~Jaeckel and S.~Schenk, ``{Exploring High Multiplicity Amplitudes in Quantum
  Mechanics},'' \href{http://dx.doi.org/10.1103/PhysRevD.98.096007}{Phys. Rev.
  D {\bfseries 98} no.~9, (2018) 096007},
  \href{http://arxiv.org/abs/1806.01857}{[arXiv:1806.01857 [hep-ph]]}.

\bibitem{Jaeckel:2018tdj}
J.~Jaeckel and S.~Schenk, ``{Exploring high multiplicity amplitudes: The
  quantum mechanics analogue of the spontaneously broken case},''
  \href{http://dx.doi.org/10.1103/PhysRevD.99.056010}{Phys. Rev. D {\bfseries
  99} no.~5, (2019) 056010},
  \href{http://arxiv.org/abs/1811.12116}{[arXiv:1811.12116 [hep-ph]]}.

\bibitem{Schenk:2019kmx}
S.~Schenk, \href{http://dx.doi.org/10.11588/heidok.00027346}{{\em {How Many
  Higgs Bosons Does it Take: Consistency of Scalar Field Theories at High
  Energies.}}}
\newblock PhD thesis, U. Heidelberg, 2019.

\bibitem{Gorsky:1993ix}
A.~S. Gorsky and M.~B. Voloshin, ``{Nonperturbative production of multiboson
  states and quantum bubbles},''
  \href{http://dx.doi.org/10.1103/PhysRevD.48.3843}{Phys. Rev. D {\bfseries 48}
  (1993) 3843--3851},
  \href{http://arxiv.org/abs/hep-ph/9305219}{[arXiv:hep-ph/9305219]}.

\bibitem{Khoze:2018kkz}
V.~V. Khoze, ``{Semiclassical computation of quantum effects in multiparticle
  production at large lambda n},''
  \href{http://arxiv.org/abs/1806.05648}{arXiv:1806.05648 [hep-ph]}.

\bibitem{Dine:2020ybn}
M.~Dine, H.~H. Patel, and J.~F. Ulbricht, ``{Behavior of Cross Sections for
  Large Numbers of Particles},''
  \href{http://arxiv.org/abs/2002.12449}{arXiv:2002.12449 [hep-ph]}.

\bibitem{Voloshin:2017flq}
M.~B. Voloshin, ``{Loops with heavy particles in production amplitudes for
  multiple Higgs bosons},''
  \href{http://dx.doi.org/10.1103/PhysRevD.95.113003}{Phys. Rev. D {\bfseries
  95} no.~11, (2017) 113003},
  \href{http://arxiv.org/abs/1704.07320}{[arXiv:1704.07320 [hep-ph]]}.

\bibitem{Libanov:1996vq}
M.~V. Libanov, ``{Multiparticle threshold amplitudes exponentiate in arbitrary
  scalar theories},'' \href{http://dx.doi.org/10.1142/S021773239600254X}{Mod.
  Phys. Lett. A {\bfseries 11} (1996) 2539--2546},
  \href{http://arxiv.org/abs/hep-th/9610036}{[arXiv:hep-th/9610036]}.

\bibitem{Goldberg:1990ys}
H.~Goldberg and M.~T. Vaughn, ``{Tree and nontree multiparticle amplitudes},''
  \href{http://dx.doi.org/10.1103/PhysRevLett.66.1267}{Phys. Rev. Lett.
  {\bfseries 66} (1991) 1267--1270}.

\bibitem{Khoze:2014kka}
V.~V. Khoze, ``{Perturbative growth of high-multiplicity W, Z and Higgs
  production processes at high energies},''
  \href{http://dx.doi.org/10.1007/JHEP03(2015)038}{JHEP {\bfseries 03} (2015)
  038}, \href{http://arxiv.org/abs/1411.2925}{[arXiv:1411.2925 [hep-ph]]}.

\bibitem{Voloshin:1992xb}
M.~B. Voloshin, ``{Zeros of tree level amplitudes at multiboson thresholds},''
  \href{http://dx.doi.org/10.1103/PhysRevD.47.2573}{Phys. Rev. D {\bfseries 47}
  (1993) 2573--2577},
  \href{http://arxiv.org/abs/hep-ph/9210244}{[arXiv:hep-ph/9210244]}.

\bibitem{Voloshin:1992nm}
M.~B. Voloshin, ``{Some properties of amplitudes at multiboson thresholds in
  spontaneously broken scalar theory},''
  \href{http://dx.doi.org/10.1103/PhysRevD.47.3525}{Phys. Rev. D {\bfseries 47}
  (1993) 3525--3529},
  \href{http://arxiv.org/abs/hep-ph/9211242}{[arXiv:hep-ph/9211242]}.

\bibitem{Argyres:1992un}
E.~N. Argyres, C.~G. Papadopoulos, and R.~H.~P. Kleiss, ``{On amplitude zeros
  at threshold},'' \href{http://dx.doi.org/10.1016/0370-2693(93)90637-W}{Phys.
  Lett. B {\bfseries 302} (1993) 70--73},
  \href{http://arxiv.org/abs/hep-ph/9212280}{[arXiv:hep-ph/9212280]}.
  [Addendum: Phys.Lett.B 319, 544 (1993)].

\bibitem{Argyres:1993xa}
E.~N. Argyres, R.~H.~P. Kleiss, and C.~G. Papadopoulos, ``{Nullification of
  multi - Higgs threshold amplitudes in the Standard Model},''
  \href{http://dx.doi.org/10.1016/0370-2693(93)91291-T}{Phys. Lett. B
  {\bfseries 308} (1993) 315--321},
  \href{http://arxiv.org/abs/hep-ph/9303322}{[arXiv:hep-ph/9303322]}.
  [Addendum: Phys.Lett.B 319, 544 (1993)].

\bibitem{Smith:1993hz}
B.~H. Smith, ``{Effects of amplitude nullification in the standard model},''
  \href{http://dx.doi.org/10.1103/PhysRevD.49.1081}{Phys. Rev. D {\bfseries 49}
  (1994) 1081--1085},
  \href{http://arxiv.org/abs/hep-ph/9309231}{[arXiv:hep-ph/9309231]}.

\bibitem{Argyres:1993wz}
E.~N. Argyres, R.~H.~P. Kleiss, and C.~G. Papadopoulos, ``{Multiscalar
  amplitudes to all orders in perturbation theory},''
  \href{http://dx.doi.org/10.1016/0370-2693(93)91287-W}{Phys. Lett. B
  {\bfseries 308} (1993) 292--296},
  \href{http://arxiv.org/abs/hep-ph/9303321}{[arXiv:hep-ph/9303321]}.
  [Addendum: Phys.Lett.B 319, 544 (1993)].

\end{thebibliography}\endgroup

\end{document}